# Mottness in a doped organic superconductor

H. Oike[1,*], K. Miyagawa[1], H. Taniguchi[2], K. Kanoda[1]
[1]*Department of Applied Physics, University of Tokyo, Bunkyo-ku, Tokyo 113-0032, Japan*
[2]*Department of Physics, Saitama University, Saitama, Saitama 338-8570, Japan*
*Current addresses: RIKEN Center for Emergent Matter Science (CEMS), Wako, Saitama 351-0198, Japan.

We report the pressure study of a doped organic superconductor with Hall coefficient and conductivity measurements. We find that maximally enhanced superconductivity and a non-Fermi liquid appear around a certain pressure where mobile carriers increase critically, suggesting a possible quantum phase transition between strongly and weakly correlated regimes. Our description extends the conventional picture of a Mott metal-insulator transition at half filling to the case of a doped Mott insulator with tunable correlation.

The interaction strength and band filling are the parameters controlling electronic phases in strongly correlated electron systems, as depicted in Fig.1(a). For repulsively interacting electrons of the same number as the lattice sites (half-filled band), the on-site Coulomb repulsion $U$, when exceeding the kinetic energy characterized by the bandwidth $W$, prevents electrons from doubly-occupying a site, thus causing them to localize at each site. This interaction-induced insulator is called a Mott insulator. Decreasing $U$ (or increasing $W$) by pressure along the blue line in Fig.1(a) causes the first-order Mott transition from the Mott insulator to a Fermi liquid or a superconductor at the Mott boundary $(U/W)_c$ [1], where a discontinuous increase in double occupancy occurs; roughly speaking, electrons are allowed to doubly-occupy a site. If the band filling is varied from a half by removing electrons from the Mott insulator, the doped holes give rise to fascinating phenomena such as high-$T_c$ superconductivity [2], non-Fermi liquid [3, 4], pseudogap [5], and self-organization of nanostructure [6-8]. However, questions remain regarding the case in which $U/W$ is varied under doping because a doped conductor with tunable $U/W$ across the critical value have not been available. The present study tackles this issue experimentally with an organic conductor, the highly compressible nature of which permits a wide range of variation in $U/W$.

The family of layered organic conductors, $\kappa$-(ET)$_2$X, with half-filled bands is well recognized as model systems for Mott physics under $U/W$ control [9-12], where ET denotes bis(ethylenedithio)tetrathiafulvalene. Pressure experiments for these conductors have demonstrated a first-order phase transition from a Mott insulator to a Fermi liquid (FL) or a superconductor [13, 14] and have revealed its criticality [15]. The purple arrow in Fig.1(a) exemplifies a range of the pressure study for $\kappa$-(ET)$_2$Cu$_2$(CN)$_3$ ($\kappa$-Cu$_2$(CN)$_3$). While most $\kappa$-ET compounds have half-filled bands, the title compound $\kappa$-(ET)$_4$Hg$_{2.89}$Br$_8$ ($\kappa$-HgBr) with the nonstoichiometry in the Hg composition [16, 17] is an exceptional doped system and shows the non-monotonic pressure dependence of superconducting (SC) transition temperature $T_c$ [18] and non-Fermi-liquid (NFL)-like resistivity [19] unlike the case of the half-filled $\kappa$-(ET)$_2$X. Assuming that the valences of Hg and Br ions are +2 and -1, respectively, the valence of an ET dimer is +1.11 due to the nonstoichiometry, while that in half-filled $\kappa$-(ET)$_2$X is +1 where X is monovalent anion. Indeed, Raman spectroscopy has confirmed that the valence of the dimer deviates from unity [20]. According to the estimate of $U/W$ based on the band-structure calculations, as described later in detail, $\kappa$-HgBr has a much larger value of $U/W$ than those of half-filled Mott insulators such as $\kappa$-Cu$_2$(CN)$_3$ [21-23] and should be in a strongly correlated state, where electrons remain considerably prevented from the double occupancy while the 11% hole doping makes the system metallic as located in Fig. 1(a) of the band filling-$U/W$ phase diagram. Thus, the pressure study of $\kappa$-HgBr affords a chance to draw together two physics regimes under the variations of the correlation strength and band filling. With the aim of investigating how the doped system with strong correlation behaves as the prohibited double occupancy is allowed, we examine the nature of mobile carriers under pressure variation and characterize the normal-state transport properties and superconductivity in a pressure-temperature diagram.

To characterize the pressure dependence of the nature of mobile carriers in $\kappa$-HgBr, we measured the Hall coefficient with four probe technique as employed in ref. [24] In order to eliminate the voltage drop due to the resistivity along current direction, the Hall voltage is determined under the reversal of field direction. We confirmed that it linearly depended on magnetic field below 5 T. The sample was put in a pressure cell, which is a dual structured clamp-type cell formed by BeCu and NiCrAl cylinders with Daphne7373 oil as a pressure medium. The normal and SC states under pressure are characterized by contactless conductivity measurements, which utilize the technique of AC susceptibility measurements in the MHz frequency range. This method probes the resistivity in the normal state because the eddy current due to electromagnetic induction causes a diamagnetic response, which yields the characteristic length of flux penetration, namely the skin depth, $\delta$ (the so-called skin effect). Because $\delta$ depends on the resistivity, an analysis of the diamagnetic response enables us to evaluate the resistivity [25]. This method is superior to the conventional four-terminal method for the present study in that a single experimental run for an identical crystal yields information on both the normal-state transport and the SC diamagnetism probing the SC volume fraction. We used a LC resonance circuit as employed in ref. [25], where $L$ is the inductance of a coil and $C$ is the capacitance of a capacitor. The sample was mounted in the coil, which was put in the dual structured clamp-type pressure cell. In the contactless conductivity measurement, we used the Daphne7474 oil, which does not solidify up to 3.7GPa [26], as a pressure-transmitting medium. The capacitance was put out of the pressure cell. AC field was applied perpendicular to the conducting plane of the sample. In this arrangement, $\chi_{rf}'$ probes the in-plane resistivity $\rho_{//}$. Because the $L$ is a function of $\chi_{rf}'$, $\rho_{//}$ is known by measuring the resonance frequency of the circuit which is determined by $L$ and $C$. Details of analysis are described in supplemental material. The single crystals of $\kappa$-HgBr used herein were grown by standard electrochemical methods.

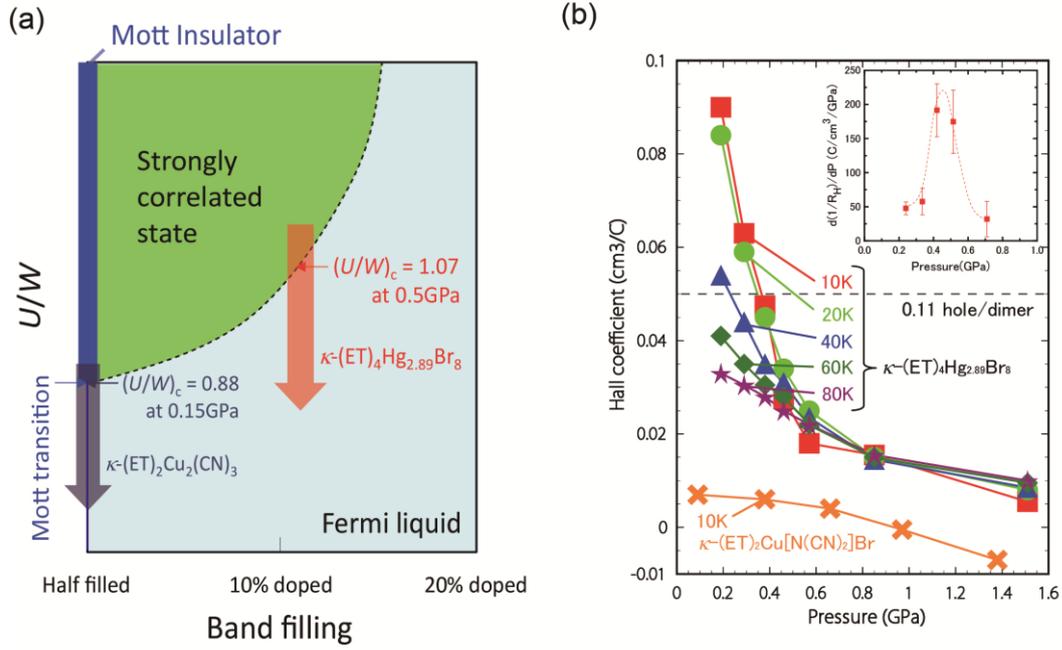

**Figure 1** Generic $U/W$-carrier density phase diagram. (a) Schematic band filling-$U/W$ phase diagram based on experimental results for $\kappa$-ET compounds. The double occupancy is considered to become allowed at the critical value in the doped organic conductor $\kappa$-(ET)$_4$Hg$_{2.89}$Br$_8$ under pressure as well as in the undoped systems. The critical value of $U/W$ increases with doping. (b) Pressure dependence of $R_H$ for $\kappa$-(ET)$_4$Hg$_{2.89}$Br$_8$ at several temperatures (the present study) and for the half-filled metallic system $\kappa$-(ET)$_2$Cu[N(CN)$_2$]Br at 10K [24]. The inset shows pressure derivative of $1/R_H$ for $\kappa$-(ET)$_4$Hg$_{2.89}$Br$_8$ at 10K. The drastic pressure dependence suggests that 0.5 GPa is a critical pressure for electrons occupying a site

The pressure dependence of the Hall coefficient $R_H$ becomes more remarkable at lower temperatures (Fig. 1(b)). At 10 K, $R_H$ steeply decreases for pressures up to 0.5 GPa and suddenly turns to leveling off. Note that the drastic change in $R_H$ occurs in a metallic state, which is reported to be stable [16, 18, 19]. Above 0.6 GPa, both magnitude and pressure dependences of $R_H$ are similar to those of the half-filled metallic system $\kappa$-(ET)$_2$Cu[N(CN)$_2$]Br [24], the $R_H$ values of which well correspond to the cross-sectional area of Fermi surfaces as in other $\kappa$-ET compounds in a conventional metallic state [24, 27]. Therefore, $\kappa$-HgBr under pressures above 0.6 GPa is considered to be in a conventional metallic state with a large Fermi surface as in $\kappa$-(ET)$_2$Cu[N(CN)$_2$]Br. However, the enormously enhanced $R_H$ values of $\kappa$-HgBr in the low-pressure range cannot be understood within the framework of the conventional metal in a weakly correlated regime because band filling is not likely to change under pressure as well as half-filled $\kappa$-ET compounds. Strong electronic correlation at ambient or low pressures is indicated by significantly enhanced electronic specific heat coefficient and nuclear magnetic resonance (NMR) relaxation rate [28, 29] and thus is likely responsible for the large $R_H$ values in the low pressure. There are several ways to interpret the enhancement of $R_H$, as proposed theoretically in a strongly correlated regime [30-32]. A simple and widely argued scenario is that the prohibition of double occupancy decreases the density of mobile carriers. Simply assuming that only doped carriers are mobile, the mobile carrier density $n$ equals to 0.11 per site, which corresponds to $5.0 \times 10^{-2}$ cm$^3$/C in $1/ne$, where $e$ is the elementary charge. The values of $R_H$ below 0.4 GPa are of the order of this value. Another scenario is that anisotropic carrier conduction due to the enhanced spin fluctuations differentiates $R_H$ from the value of the conventional metal as argued in the fluctuation-exchange theory [31] and $t$-$t'$-$J$ model [32]. Considering that the spin fluctuations are enhanced due to the prohibition of double occupancy, both scenarios suggest that there occurs an anomaly in double occupancy around 0.5 GPa. The compressibility of $1/R_H$, which measures the density of mobile carriers, exhibits a sharp peak near 0.5 GPa (see inset of Fig. 1). Thus, 0.5 GPa is considered to be a critical pressure for the double occupancy, pointing to a sharp change from a strongly correlated state to a FL. (see also Fig.1(a)).

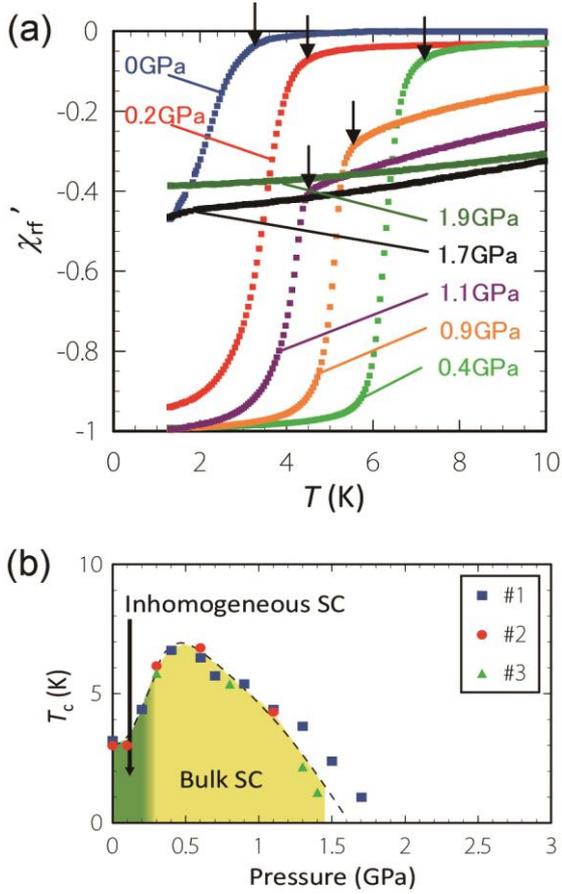

**Figure 2** Superconducting transition probed by AC susceptibility and pressure dependence of the transition temperature. (a) $\chi_{rf}'$ of $\kappa$-(ET)$_4$Hg$_{2.89}$Br$_8$ under pressure. The arrows indicate SC transitions. The value of $\chi_{rf}'$ at pressures between 0.3 GPa and 1.3 GPa sharply saturates to the same value, which is interpreted as perfect diamagnetism. (b) Pressure dependence of $T_c$ for the three samples investigated, #1, #2 and #3. The SC transitions occur in the pressure range of 0-1.5 GPa. At ambient pressure and 0.2 GPa, the SC diamagnetism is not perfect.

SC diamagnetic responses are observed below the temperatures indicated by arrows in Fig. 2(a) and are imposed on eddy-current-induced diamagnetism in the normal state (discussed below). At ambient pressure and 0.2 GPa, the absolute value of $\chi_{rf}'$ extrapolated to 0 K did not reach the value of perfect diamagnetism, suggesting that both SC and non-SC regions coexist, where $\chi_{rf}'$ is the real part of the AC susceptibility. At 0.4 GPa and higher, however, the sample becomes fully superconducting. We confirmed the dome structure of $T_c$ near 0.5 GPa (Fig. 2(b)). A SC transition was not observed above 1.7 GPa in the present study, which indicates that the SC transition observed above this pressure in the previous four-probe resistivity measurement [19] was not a bulk transition. Because the spatial inhomogeneity of SC at 0.2 GPa and lower is eliminated by the pressure, the emergence of inhomogeneity is likely inherent of the strongly correlated regime.

In the normal state, diamagnetic responses due to the skin effect were observed, and the in-plane resistivity $\rho_{//}$ was obtained from the analysis of AC susceptibility $\chi_{rf}'$. Paramagnetic contribution, which is the order of $10^{-5}$ in $\kappa$-HgBr [28], is negligibly small in comparison to $\chi_{rf}'$, which is the order of $10^{-1}$ in the present measurements. As described in detail in supplementary information, the $\rho_{//}$ values are reliable when skin depth $\delta$ is shorter than sample size $r$. We checked the reproducibility of their pressure and temperature dependence although absolute values $\rho_{//}$ are different from that measured in other samples by a factor of 3. Figure 3(a) shows $\rho_{//}$ for several pressures. At a low pressure of 0.3 GPa, $\rho_{//}$ exhibits a convex curve as a function of temperature. At intermediate pressures, 0.6 GPa and 1.1 GPa, $\rho_{//}$ exhibits a linear temperature dependence down to $T_c$, clearly indicating the NFL behavior that persists to $T_c$. At 1.4 GPa and higher, $\rho_{//}$ exhibits concave curves at low temperatures, where $\rho_{//}$ is well approximated by a form of $\rho_{//} = \rho_o + AT^2$. However, this behavior appears to cross over to a linear temperature dependence at higher temperatures (Fig. 3(b)). To further examine the FL and NFL regions in the pressure-temperature phase diagram, we performed an analysis to determine the "local" exponent, $\alpha$, which is defined by $\alpha = d(\log(\rho_{//}-\rho_o))/d(\log(T))$, where $\rho_o$ is the residual resistivity determined by fitting the form of $\rho_{//}-\rho_o \sim T^\alpha$ to the resistivity data below 15 K. The values of $\alpha$ are represented by a range of colors in the temperature-pressure plane in Fig. 3(c). The unexpected exponent of $\alpha < 1$ (corresponding to the concave curve) in the red-colored region well below 0.5 GPa likely reflects an inhomogeneous state leading to the imperfect SC discussed above. NMR line broadening observed below 40 K at ambient pressure is also an indication of inhomogeneity [28]. In the presence of strong electron correlation, doping is argued to cause spatially inhomogeneous phases because of their energetically competing electronic states [33-35], where disorder may work for the appearance of the inhomogeneity through spatially pinning and/or amplifying the inhomogeneity. The inhomogeneity in $\kappa$-HgBr may be a hallmark of a strongly correlated state. In between the red-colored region and the blue-colored FL regions, a NFL region of $\alpha \sim 1$ appears and becomes confined between 0.5 GPa and 1.0 GPa at low temperatures. The linear-temperature dependence of resistivity is a hallmark of non-Fermi liquid nature as observed in heavy electron systems, where the behavior appears around a quantum critical point separating the magnetically ordered phase and the heavy-electron state. Considering that the critical pressure of the double occupancy probed by the Hall coefficient falls in this range, the NFL behavior is most likely a manifestation of the critical fluctuations between the strongly correlated state and the FL.

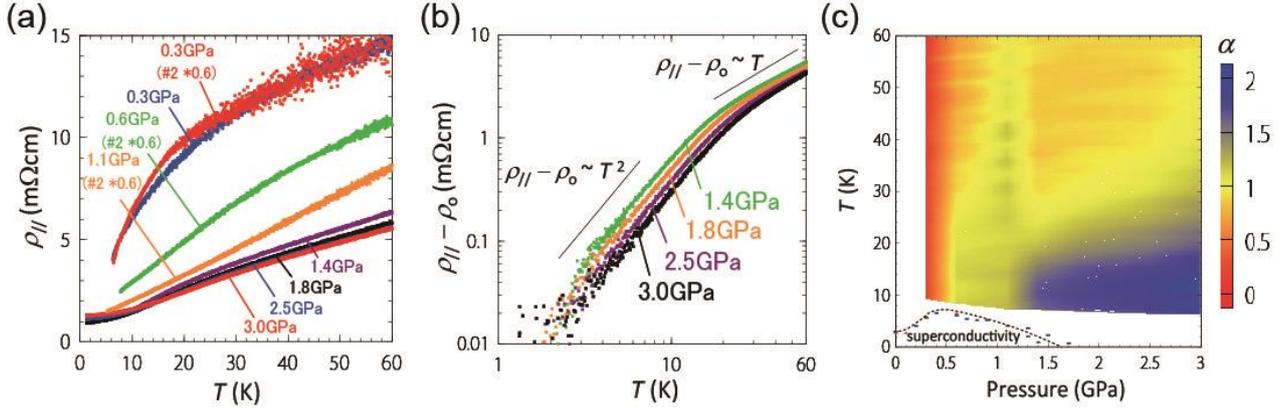

**Figure 3** Temperature dependence of normal-state resistivity under pressures. (a) $\rho_{//}$ of $\kappa$-(ET)$_4$Hg$_{2.89}$Br$_8$ obtained from the analysis of $\chi_{\rm rf}'$. (b) Temperature dependence of $\rho_{//}-\rho_o$ in a pressure range of 1.4-3.0 GPa on a logarithmic scale. (c) Contour plot of $\alpha$ in $\rho_{//}-\rho_o \sim T^\alpha$. At pressures below 0.3 GPa, the diamagnetic response due to the skin effect was too small to detect because the large value of $\rho_{//}$ increased the skin depth.

The experimental results are summarized in the phase diagram in Fig. 4, which, together with Fig. 3(c), indicates that as $U/W$ decreases, the inhomogeneous metal transforms into a FL through a NFL, the region of which becomes narrower in pressure to reside near the top of the SC dome at low temperatures. The critical pressure of 0.5 GPa for the double occupancy lies in this region. These features suggest that the crossover from the strongly correlated state to the FL may be sharpened into a quantum phase transition near the critical value, $(U/W)_c$. This is regarded as a generalization of Mott transition into a non-half-filled case in that the metal-to-metal transition (or sharp crossover) in the present doped case and the insulator-to metal transition in the non-doped case are both associated with a drastic change in Mottness, namely the degree of double occupancy. We mention the possible effect of disorder. Although it only causes to increase the residual resistivity in a Fermi liquid, it can be more significant in a critical region; that is, the non-negligible disorder indicated by the relatively large residual resistivity (~1 mΩcm) comparable to the Mott-Ioffe-Regel limit, most likely due to anion nonstoichiometry, may render the intrinsic phase transition less sharp [36]; a quantum transition or a weak first-order transition as theoretically predicted in a clean limit [37] can become crossover-like in reality. However, the anomalies around 0.5GPa do not originate from disorder-induced phase transition, such as Anderson localization, which contradicts the maximum in the pressure dependence of $T_c$ at 0.5GPa and the finite carrier density below 0.5GPa indicated by $1/R_H$, whose temperature dependence is shown in supplementary information, although there possibly exists an insulating phase in a hypothetical negative pressure range. Whether quantum phase transition exists in doped systems has been an intensively debated issue related to the high-$T_c$ superconductivity [4, 38-44]. The present study shows that a quantum phase transition or a sharp crossover accompanies SC dome under variation of correlation strength.

The case for the critical value, $(U/W)_c$, in the present doped system should correspond to the Mott transition in the half-filled case. Thus, we compare $(U/W)_c$ of $\kappa$-HgBr with that of a half-filled system with a similar lattice geometry, $\kappa$-Cu$_2$(CN)$_3$, which exhibits a Mott transition at 0.15 GPa, to extend our discussion toward a comprehensive understanding of the band filling-$U/W$ phase diagram (Fig.1(a)). The incommensurate structure of $\kappa$-HgBr makes the first-principle calculation difficult, so we employed the calculations based on extended Huckel and tight binding approximations. Although the absolute value of $U/W$ depends on the methods of the calculations [21, 45], it is meaningful to compare the values calculated by the same method. At ambient pressure, the $U/W$ values are 1.1 in $\kappa$-HgBr and 0.9 in $\kappa$-Cu$_2$(CN)$_3$, according to ref. [21]. Assuming that $U/W$ decreases at a rate of approximately 4%/GPa for both compounds similarly to the case of $\kappa$-(ET)$_2$Cu(NCS)$_2$ [46], $(U/W)_c$ is estimated to be 1.07 for $\kappa$-HgBr and 0.88 for $\kappa$-Cu$_2$(CN)$_3$. Referring to these values, the doped and undoped compounds under pressure variation are located in the band filling-$U/W$ phase diagram, as shown in Fig. 1(a). When moving from $\kappa$-Cu$_2$(CN)$_3$ to $\kappa$-HgBr, in the diagram, the $(U/W)_c$ value is increased possibly due to doping. The mobile carriers generated by the doping enhance the screening effect and weakens the effective interaction, which explains the increase in $(U/W)_c$.

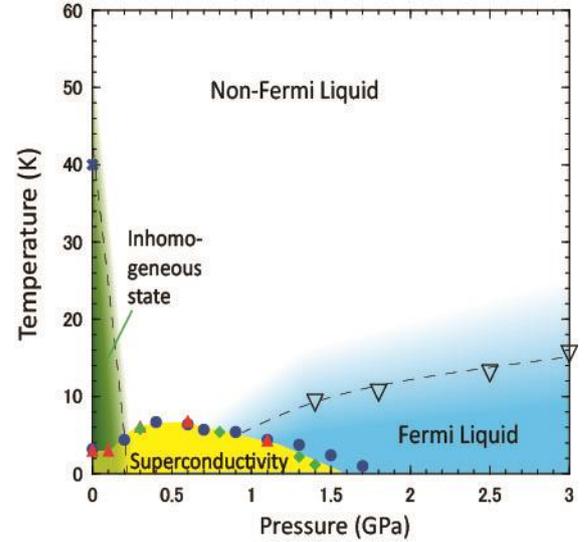

**Figure 4** Pressure-temperature phase diagram for $\kappa$-(ET)$_4$Hg$_{2.89}$Br$_8$. Pressure-temperature phase diagram of $\kappa$-(ET)$_4$Hg$_{2.89}$Br$_8$. At ambient pressure and 0.2 GPa, the SC is inhomogeneous. Likewise, the $^{13}$C NMR line is broadened below 40 K at ambient pressure [28], indicating an inhomogeneous state. These signatures are reflected by the green color. The open triangles indicate the temperatures where the resistivity starts to deviate from the Fermi liquid behavior of $\rho_{//}= \rho_o + AT^2$. The dashed lines serve as a guide to the eye.

In conclusion, the pressure study of an organic superconductor with a band filling away from a half revealed that the top of the superconducting dome, the appearance of non-Fermi liquid behavior and a change in the electronic homogeneity all occur around a certain value of $U/W$ where the density of mobile carriers shows a critical increase. Viewing the results in the band filling-$U/W$ phase diagram, the present observation is addressed as an extension of the conventional Mott transition to the doped case, where the criticality of Mottness resides at a finite level of doping.

**Acknowledgements** We thank K. Murata for useful suggestions on high-pressure techniques. This work was supported in part by JSPS KAKENHI under Grant Nos. 20110002, 25220709, 24654101, and 11J09324 and the US National Science Foundation under Grant No. PHYS-1066293 and the hospitality of the Aspen Center for Physics.

**References and Notes:**
1. M. Imada, A. Fujimori, and Y. Tokura, Rev. Mod. Phys. **70**, 1039 (1998).
2. J. G. Bednorz, and K. A. Müller, Z. Phys. B **64**, 189 (1986).
3. H. Takagi, B. Batlogg, H. L. Kao, J. Kwo, R. J. Cava, J. J. Krajewski, and W. F. Peck, Jr., Phys. Rev. Lett. **69**, 2975 (1992).
4. R. A. Cooper, Y. Wang, B. Vignolle, O. J. Lipscombe, S. M. Hayden, Y. Tanabe, T. Adachi, Y. Koike, M. Nohara, H. Takagi, Cyril Proust, and N. E. Hussey, Science **323**, 603 (2009).
5. R. Liu, B. W. Veal, A. P. Paulikas, J. W. Downey, P. J. Kostić, S. Fleshler, U. Welp, C. G. Olson, X. Wu, A. J. Arko, and J. J. Joyce, Phys. Rev. B **46**, 11056 (1992).
6. K. M. Lang, V. Madhavan, J. E. Hoffman, E. W. Hudson, H. Eisaki, S. Uchida, and J. C. Davis, Nature **415**, 412 (2002).
7. E. Dagotto, Science **309**, 257 (2005).
8. K. M. Lang, V. Madhavan, J. E. Hoffman, E. W. Hudson, H. Eisaki, S. Uchida and J. C. Davis, Science **315**, 1380 (2007).
9. K. Kanoda, Physica C **282-287**, 299 (1997).
10. K. Kanoda, Hyperfine Interactions **104**, 235 (1997).
11. K. Miyagawa, K. Kanoda, and A. Kawamoto, Chem. Rev. **104**, 5635 (2004).
12. B. J. Powell, and R. H. McKenzie, Rep. Prog. Phys. **74**, 056501 (2011)
13. S. Lefebvre, P. Wzietek, S. Brown, C. Bourbonnais, D. Jérome, C. Mézière, M. Fourmigué, and P. Batail, Phys. Rev. Lett. **85**, 5420 (2000).
14. P. Limelette, P.Wzietek, S. Florens, A. Georges, T. A. Costi, C. Pasquier, D. Je´rome, C. Me´zie`re, and P. Batail, Phys. Rev. Lett. **91**, 016401 (2003).
15. F. Kagawa, K. Miyagawa, K. Kanoda, Nature **436**, 534 (2005).
16. R. N. Lyubovskaya, E. A. Zhilyaeva, A. V. Zvarykina, V. N. Laukhin, R. B. Lyubovskii, and S. I. Pesotskii, JETP Lett. **45**, 530 (1987).
17. R. N. Lyubovskaya, E. I. Zhilyaeva, S. I. Pesotskii, R. B. Lyubovskii, L. O. Atovmyan, O. A. D'yachenko, and T. G. Takhirov, JETP Lett. **46**, 188 (1987).
18. S. L. Bud'ko, A. G. Gapotchenko, A. E. Luppov, R. N. Lyubovskaya, and R. B. Lyubovskii, Sov. Phys. JETP **74**, 983 (1992).
19. H. Taniguchi, T. Okuhata, T. Nagai, K. Satoh, N. Mori, Y. Shimizu, M. Hedo, and Y. Uwatoko, J. Phys. Soc. Jpn. **76**, 113709 (2007).
20. T. Yamamoto, , M. Uruichi, , K. Yamamoto, , K. Yakushi, , A. Kawamoto, and H. Taniguchi, J. Phys. Chem. B **109**, 15226 (2005).
21. T. Mori, A. Kobayashi, Y. Sasaki, H. Kobayashi, and H. Inokuchi, Bull. Chem. Soc. Jpn. **57**, 627 (1984).
22. U. Geiser , H. H. Wang , K. D. Carlson, J. M. Williams, H. A. Charlier, J. E. Heindl, G. A. Yaconi, and B. J. Love, Inorganic Chemistry, **30**, 2586 (1991).
23. R. Li, V. Petricek, G. Yang, P. Coppens, and M. Naughton, Chem. Mater. **10**, 1521 (1998).
24. K. Katayama, T. Nagai, H. Taniguchi, K. Satoh, N. Tajima, and R. Kato, J. Low. Temp. Phys. **142**, 515 (2006).
25. H. Oike, K. Miyagawa, K. Kanoda, H. Taniguchi, and K. Murata, Physica B **404**, 376 (2009).
26. K. Murata, K. Yokogawa, H. Yoshino, S. Klotz, P. Munsch, A. Irizawa, M. Nishiyama, K. Iizuka, T. Nanba, T. Okada, Y. Shiraga, and S. Aoyama, Rev. Sci. Instrum. **79**, 085101 (2008).
27. T. Sasaki, and N. Toyota, Synth. Met. **55-57**, 2303 (1993).
28. Y. Eto, M. Itaya, and A. Kawamoto, Phys. Rev. B **81**, 212503 (2010).
29. A. Naito, Y. Nakazawa, K. Saito, H. Taniguchi, K. Kanoda, and M. Sorai, Phys. Rev. B **71**, 054514 (2005)

30. E. Lange, and G. Kotliar, Phys. Rev. B **59**, 1800 (1999).
31. H. Kontani, K. Kanki, and K. Ueda, Phys. Rev. B **59**, 14732 (1999).
32. J. O. Haerter, and B. S. Shastry, Phys. Rev. B **77**, 045127 (1999).
33. V. J. Emery, and S. A. Kivelson, Physica C **209**, 597 (1993).
34. J. Schmalian, and P. G. Wolynes Stripe, Phys. Rev. Lett. **85**, 836 (2000).
35. A. Himeda, T. Kato, and M. Ogata, Phys. Rev. Lett. **88**, 117001 (2002).
36. E. C. Andrade, E. Miranda, and V. Dobrosavljevic´, Phys. Rev. Lett. **102**, 206403 (2009).
37. G. Sordi, K. Haule, and A. M. S. Tremblay, Phys. Rev. Lett. **104**, 226402 (2010).
38. M. Capone, M. Fabrizio, C. Castellani, and E. Tosatti, Phys. Rev. Lett. **93**, 047001 (2004)
39. J. Zaanen, and B. J. Overbosch, Phil. Trans. R. Soc. A **369**, 1599 (2011).
40. Y. Yamaji, and M. Imada, Phys. Rev. Lett. **106**, 016404 (2011).
41. G. Sordi, P. Semon, K. Haule, and A. M. S. Tremblay, Phys. Rev. Lett. **108**, 216401 (2012).
42. E. Gull, O. Parcollet, and A. J. Millis, Phys. Rev. Lett. **110**, 216405 (2013).
43. H. Yokoyama, M. Ogata, Y. Tanaka, K. Kobayashi, and K. Tsuchiura, J. Phys. Soc. Jpn. **82**, 014707 (2013).
44. L. F. Tocchio, H. Lee, H. O. Jeschke, R. Valenti, and C. Gros, Phys. Rev. B **87**, 045111 (2013)
45. H. C. Kandpal, I. Opahle, Y. Z. Zhang, H. O. Jeschke, and R. Valentı, Phys. Rev. Lett. **103**, 067004 (2009).
46. M. Rahal, D. Chasseau, J. Gaultier, L. Ducasse, M. Kurmoo, and P. Day, Acta Cryst. B **53**, 159 (1997).

# Supplementary information

## I. CONTACTLESS CONDUCTIVITY MEASUREMENT

**Principle of contactless conductivity measurement.** The resistivity of metallic samples can be determined from the AC susceptibility $\chi_{rf}'$ in a MHz-frequency range without taking electrical contacts to the sample. The $\chi_{rf}'$ of the highly conducting state comes from the frequency-independent Pauli paramagnetism and Landau diamagnetism, and a frequency-dependent diamagnetism due to the electromagnetically induced shielding current (so called as the skin effect). While the former two contributions in organic conductors are typically of the order of $10^{-6}$, the latter is estimated, e.g., at the order of $10^{-1}$ at 10K under 1.1 GPa for the present experimental conditions as described below; so the latter, overwhelming the former, determines the values $\chi_{rf}'$ in the highly conducting state. To obtain $\chi_{rf}'$, we used a LC resonance circuit as employed in ref. [1] (Fig. 1), where $L$ is the inductance of a coil and $C$ is the capacitance of a capacitor. A coil in which a single crystal of $\kappa$-(ET)$_4$Hg$_{2.89}$Br$_8$ was mounted was put in the pressure cell and connected with the capacitors set out of the cell to construct a resonant circuit, where the capacitor 1 tunes the resonance frequency, while the capacitor 2 was mainly for impedance matching. The circuit was connected to Network Analyzer (Agilent Technologies: E5061A), which can measure the frequency dependence of $\Gamma$, the radio frequency (RF) power reflected from the circuit. We measured $f_{dip}$, the frequency at which $\Gamma$ took a minimum value. Because the $L$ is a function of $\chi_{rf}'$, $\rho_{//}$ is known by measuring $f_{dip}$ which is determined by $L$ and $C$.

**Impedance matching condition.** In order to obtain $\chi_{rf}'$ from $f_{dip}$, it is necessary to know the frequency dependence of $Z$ in detail, where $Z (= R + i X)$ is the impedance of the circuit. The output and input impedances of the Network Analyzer are 50ohm, so $\Gamma$ is described as

$$\Gamma = \left| \frac{(R - 50\Omega) + iX}{(R + 50\Omega) + iX} \right|, \quad (1)$$

where imaginary part of the output impedance is regarded as zero. When $R = 50$ohm and $X = 0$ ohm, $\Gamma$ is zero (so called impedance matching). When $R \neq 50$ohm, $\Gamma$ takes a finite value. Thus, with the Network Analyzer, we searched the frequency where the impedance matching is realized.

Since $Z$ of the ideal circuit shown in Fig. 1 is purely imaginary number, it seems that $\Gamma = 1$ at any frequency according to the equation (1). In reality, however, $R$ is non-zero because of the resistances of the circuit elements. We took into account only the resistance of the coil, and neglected the resistance of the other circuit elements and dielectric loss in the capacitors (Fig. 2(a)). Then, $Z$ is expressed as

$$Z = \frac{1}{2\pi i f C_1 + \dfrac{1}{r_L + 2\pi i f L}} + \frac{1}{2\pi i f C_2}, \quad (2)$$

where $f$, $r_L$, $C_1$, and $C_2$ are the frequency of input RF wave, the resistance of the coil, the capacitances of capacitor 1 and 2, respectively. $r_L / 2\pi f L$ was the order of $10^{-3}$, so we neglected the second or higher order of this term. Then, $R$ and $X$ can be approximated as

$$R = \frac{r_L}{\{LC_1(2\pi f)^2 - 1\}^2},$$

$$X = -\frac{L(C_1 + C_2)(2\pi f)^2 - 1}{2\pi f C_2 \{LC_1(2\pi f)^2 - 1\}}. \quad (3)$$

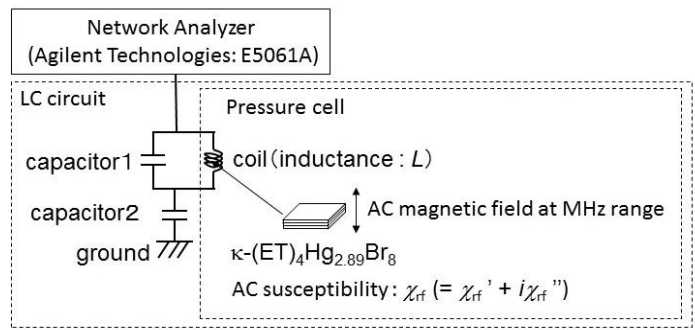

Fig. 1. Experimental setting of contactless conductivity measurement. A coil in which a single crystal of $\kappa$-(ET)$_4$Hg$_{2.89}$Br$_8$ was mounted was put in the pressure cell and connected with the capacitors set out of the cell to construct a resonant circuit, where the capacitor 1 tunes the resonance frequency, while the capacitor 2 was mainly used for impedance matching. The circuit was connected to Network Analyzer (Agilent Technologies: E5061A).

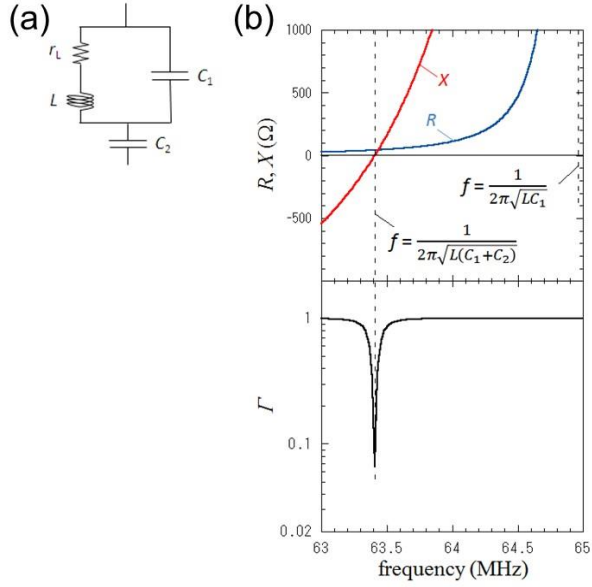

Fig. 2. Impedance matching condition. (a) Equivalent circuit of the LC circuit. (b) Frequency dependence of $R$ and $X$ of the equivalent circuit. Circuit constants used in this calculation are described in text of supplementary materials. $\Gamma$ takes minimum at the frequency $f = [L(C_1+ C_2)]^{-0.5}/2\pi$, where $X$ is 0.

The profiles of frequency dependences of $R$ and $X$ are shown in Fig. 2(b) in case that $r_L$, $L$, $C_1$, and $C_2$ are 0.1ohm, 300nH, 20pF and 1pF, respectively. $R$ diverges at $2\pi f = (LC_1)^{-0.5}$ and $X$ becomes zero at $2\pi f = [L(C_1+ C_2)]^{-0.5}$. We can obtain the impedance matching condition ($R$ = 50ohm, $X$ = 0ohm) by varying $C_2$. In fact, $f_{dip}$ slightly deviated from the impedance matching condition on temperature sweep because the circuit constants are temperature-dependent. $\Gamma$ at $f_{dip}$ in this experiment was smaller than 0.2, which corresponds to the $R$ and $X$ values of 30-80ohm and -20-20ohm, respectively, according to the equation (3). When $R$ and $X$ take these values, the frequency dependence of $X$ is much steeper than that of $R$, and mainly contributes to that of $\Gamma$, consequently. Then, $\Gamma$ takes minimum when $X = 0$, and $f_{dip}$ is described as

$$f_{dip} = \frac{1}{2\pi\sqrt{L(C_1 + C_2)}}. \tag{4}$$

Thus, $f_{dip}$ was expressed as a function of $L$.

**Cancellation of extrinsic frequency change.** When a sample is put in the coil, $f_{dip}$ can be written as

$$f_{dip} = f_o + \Delta f_\chi + \Delta f_o, \tag{5}$$

where $f_o$ is a resonance frequency in case of no sample, $\Delta f_\chi$ is a frequency shift caused by the shielding diamagnetism of the sample and $\Delta f_o$ is unavoidable uncertainty due to a slight change of the circuit parameters after the sample was mounted. We confirmed that $f_{dip}$ in the absence of sample (background, BG) had no peculiar temperature dependence. Fig. 3(a) shows the temperature dependence of $f_{dip}$. For all pressures, $f_{dip}$'s are approximately the same as $f_{dip}$ in the absence of sample around 120 K, indicating that $\Delta f_\chi$ is negligibly small in such high temperatures and $\Delta f_o$ is not appreciable. To extract $\Delta f_\chi$ from $f_{dip}$, we first normalized $f_{dip}$ at a temperature $T_o$ where the resistivity is so large that the skin effect, that is $\Delta f_\chi$, is practically vanishing. From the equation (5), the normalized $f_{dip}$ is expressed as

$$\frac{f_{dip}(T)}{f_{dip}(T_o)} = \frac{f_o(T) + \Delta f_o(T)}{f_o(T_o) + \Delta f_o(T_o)} + \frac{\Delta f_\chi(T)}{f_o(T_o) + \Delta f_o(T_o)}. \tag{6}$$

Because $\Delta f_o$ is one hundredth of $f_{dip}$ or less, we eliminated the second order of $\Delta f_o / f_o$ and neglected its temperature dependence. Then, the equation (6) is reduced to

$$\frac{f_{dip}(T)}{f_{dip}(T_o)} = \frac{f_o(T)}{f_o(T_o)} + \frac{\Delta f_\chi(T)}{f_o(T_o) + \Delta f_o(T_o)}. \quad (7)$$

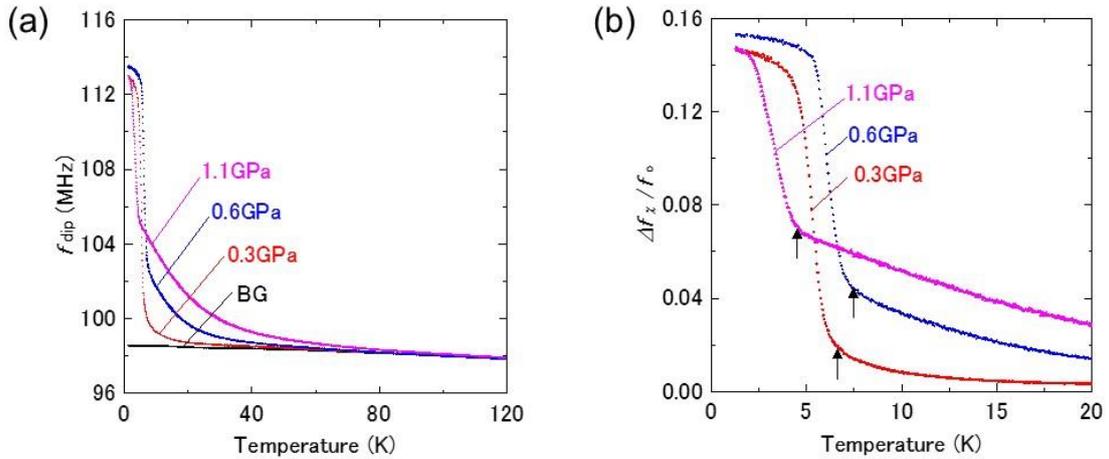

Fig. 3. Analyses to obtain a frequency shift caused by the shielding diamagnetism of the sample. Temperature dependence of $f_{dip}$ (a), and $\Delta f_\chi / f_o$ (b) for sample #2 under pressure are plotted, together with those before mounting the sample, which is represented as BG. Arrows in (b) indicate superconducting transitions.

We adopted a temperature of 190 K as $T_o$ for the measurements of sample #2, because, for all pressures measured, $f_{dip}(T)$'s nearly coincide with $f_o(T)$ around 190 K. Next, both sides of the equation (7) were divided by $f_o(T) / f_o(T_o)$ to give a form of

$$\frac{f_{dip}(T)/f_{dip}(T_o)}{f_o(T)/f_o(T_o)} = 1 + \frac{\Delta f_\chi(T)}{f_o(T)}\left(1 + \frac{\Delta f_o(T_o)}{f_o(T_o)}\right)^{-1}. \quad (8)$$

The $\Delta f_o / f_o$ is smaller than 0.01; so, by omitting it from Eq. (8), $\Delta f_\chi / f_o$ is given in a form of

$$\frac{\Delta f_\chi(T)}{f_0(T)} = \frac{f_{dip}(T)/f_{dip}(T_o)}{f_o(T)/f_o(T_o)} - 1. \quad (9)$$

**Frequency change due to diamagnetic moment.** Because $f_{dip}$ is inversely proportional to the square root of $L$, which has a linear dependence on $\chi_{rf}'$, $\Delta f_\chi$ is expressed as

$$\frac{\Delta f_\chi(T)}{f_o(T)} = \left(1 + \frac{\beta \chi_{rf}'}{1-N}\right)^{-0.5} - 1, \quad (10)$$

where $N$ is the demagnetization coefficient and $\beta$ is a constant determined by the sample setting like the filling factor of the sample in the coil in volume. Usually $\beta$ is hard to know; however, in the present experiment, it can be known from the experimental data in the superconducting state. Fig. 3(b) shows the temperature dependence of $\Delta f_\chi(T)/f_o(T)$ Under 0.3 GPa - 1.1 GPa, it steeply increases below the superconducting (SC) transition temperatures pointed by arrows in the inset and saturates to approximately the same value within the error of 3 %, pointing to the perfect diamagnetism of the sample. (Increases in $\Delta f_\chi(T)/f_o(T)$ above the SC transition temperatures come from the skin effect.) As the field penetration depth of superconductivity in $\kappa$-ET compounds is known to be of the order of μm or less [2-4], which is negligibly small compared with the sample size (~ mm), the low-temperature saturation should correspond to $\chi_{rf}' = -1$. By substituting the saturation value of $\Delta f_\chi/f_o$ determined from Eq. (9) to the left side of Eq. (10), we obtain the value of $\beta/1-N$, which is 0.25 for sample #2. Because $N$ depends on the sample shape and the volume of the flux-penetrating region, we estimated the values of $N$ under the following approximations as depicted in Fig. 4(a): (i) the shape of the sample is a spheroid with a radius of $r$ and a thickness of $2b$, and (ii) the magnetic flux are completely excluded from a reduced spheroid with a radius of $r \times (-\chi_{rf}')$ (= $r'$) and a thickness of $2b$. Then, $N$ is expressed as

$$N = \frac{1+\varepsilon^2}{\varepsilon^3}\left(\varepsilon - \tan^{-1}\varepsilon\right), \quad \varepsilon = \sqrt{1 - \left(\frac{r'}{b}\right)^2}, \quad (b > r'),$$

$$N = \frac{1-\varepsilon^2}{2\varepsilon^3}\left(\log\frac{1+\varepsilon}{1-\varepsilon} - 2\varepsilon\right), \quad \varepsilon = \sqrt{\left(\frac{r'}{b}\right)^2 - 1}, \quad (b < r'), \quad (11)$$

where $\varepsilon$ is the eccentricity of the spheroid with the radius of $r'$ and the thickness of $2b$ [5]. When the sample is in the Meissner state, $r'$ equals to $r$. Using the $r$ and $2b$ values of sample #2, which are 0.6 mm and 0.5 mm, respectively, as listed in Table 1, Eq. (11) yields $N = 0.58$ in the Meissner state. Then, we obtain the value of $\beta$ (= 0.11 for sample #2), with which the experimental value of $\Delta f_\chi(T)/f_o(T)$ is converted into $\chi_{rf}'$, taking account of the $\chi_{rf}'$ dependence of $N$ through $r'$. Although the sample is in a plate-like shape, we approximated it as a spheroid shape because exact solutions of $N$ are available in the range of $-1 \leq \chi_{rf}' \leq 0$ only for a spheroid.

**Obtaining resistivity from susceptibility.** $\chi_{rf}'$ is a function of $\rho_{//}$ through the skin effect in the normal state. To express $\chi_{rf}'$ as a function of $\rho_{//}$, we approximated the shape of the conducting plane as a circular one with a radius of $r$ and assumed the magnetic flux to penetrate into the sample from the lateral faces as depicted in Fig. 4(b). In this approximation, $\chi_{rf}$ is expressed as the susceptibility of a cylindrical conductor with an infinite length in the form of

$$\chi_{rf}' = \text{Re}\left(-1 + \frac{2}{kr}\frac{J_1(kr)}{J_0(kr)}\right),$$

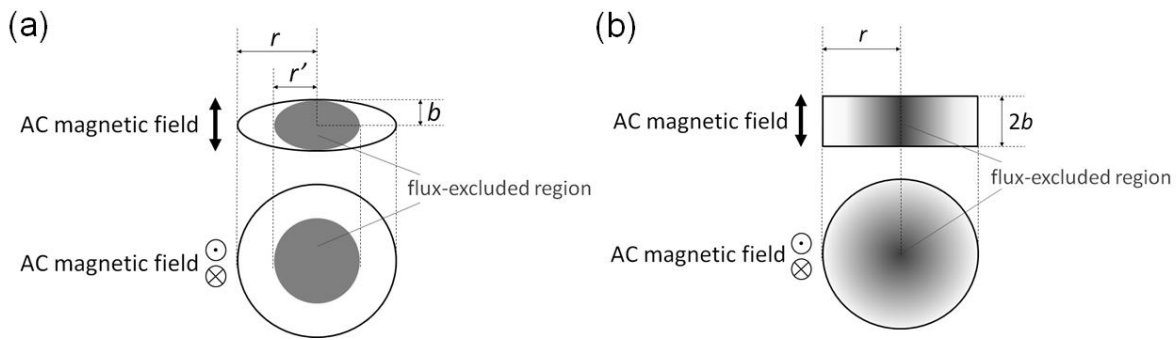

Fig. 4. Approximations of flux penetration in the sample. (a) Approximation for evaluating the demagnetization coefficient. The shapes of sample and flux-excluded region are approximated to spheroids. (b) Approximation to simplify the relationship between AC susceptibility and resistivity. The flux penetrates into the sample from the surface parallel to the applied AC magnetic field.

$$k = \frac{1+i}{\delta}, \quad \delta = \sqrt{\frac{\rho_{//}}{\pi\mu_o f}}, \quad (12)$$

with $J_n(x)$ the Bessel functions [5]. The Eq (12) shows that $\chi_{rf}'$ is a function of $\delta/r$, which is an index of how deeply the magnetic flux penetrates into the sample from the peripheral surface. Substitution of the values of $\chi_{rf}'$ to Eq. (12) yields the values of $\rho_{//}$.

**Reliability of obtained resistivity values.** The obtained values of $\rho_{//}$ in the analysis described above have ambiguity, for which two origins are conceivable. The first one is the crudeness of the approximation for the relationship between $\chi_{rf}'$ and $\rho_{//}$. In case that $\delta$ is much smaller than $r$, deviation of $\chi_{rf}'$ from the perfect diamagnetism is proportional to $\delta$. This is because the flux excluded region, which is proportional to deviation of $\chi_{rf}'$ from -1, is approximated to be $\delta \times$ (perimeter) $\times$ (thickness). This relation holds even when the shape of conducting plane is not circle. Then, the temperature and pressure dependences of $\delta$ obtained in the present analysis is justified although the estimation of (perimeter) $\times$ (thickness) affects its absolute value more or less. On the other hand, when $\delta$ is larger than $r$, the way of flux penetration depends on the shape of the sample. As a result, it is likely to have uncertainties in calculating $\rho_{//}$. The second origin is possible non-zero value of $\Box f_o$ in the equation (6). When $\Delta f_\chi/f_o$ become small enough to approach the second order and/or the temperature variation of $\Delta f_o/f_o$, which are neglected in the equations (7)-(10), the obtained values of $\chi_{rf}'$ get ambiguous.

The ambiguity due to the first and the second origins could be minimized by designing the resonance circuit so as to operate it at high frequencies and/or using a larger sample. According to the equation (12), increases of the frequency shortens $\delta$, thus decreasing the value of $\delta/r$. Then the absolute value of $\chi_{rf}'$ becomes large and consequently $\Delta f_\chi/f_o$ does. In the measurements of samples #2 and #3, we decreased the value of $\delta/r$ using the higher frequency and the larger sample than the measurements of #1 as listed in Table 1. Fig. 5 shows the temperature dependences of $\rho_{//}$ for samples #1, #2 and #3. All of three samples reproduced the metallic behavior.

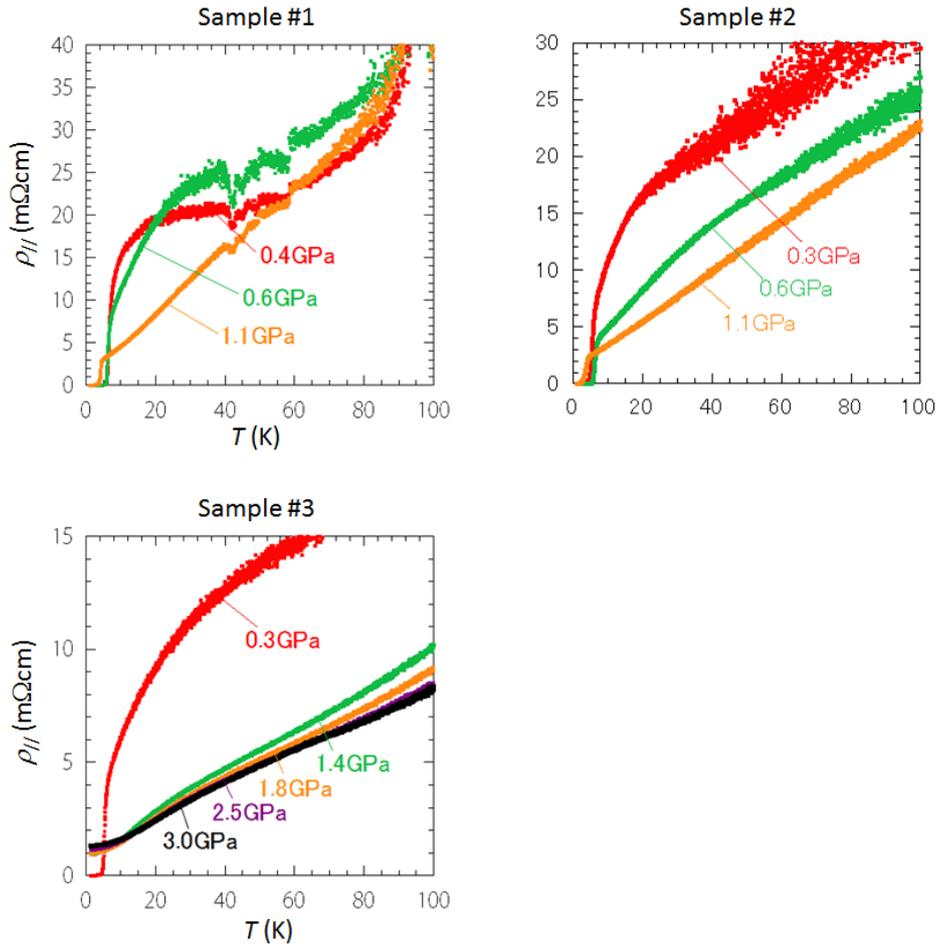

Fig. 5. Temperature dependence of $\rho_{//}$ in the measurements on three samples. All of three samples reproduced the metallic behavior, but the temperature profile of $\rho_{//}$ deviated remarkably from each other in the high temperature range due to the ambiguities in approximation to obtain $\rho_{//}$, detail of which is described in text of supplementary materials.

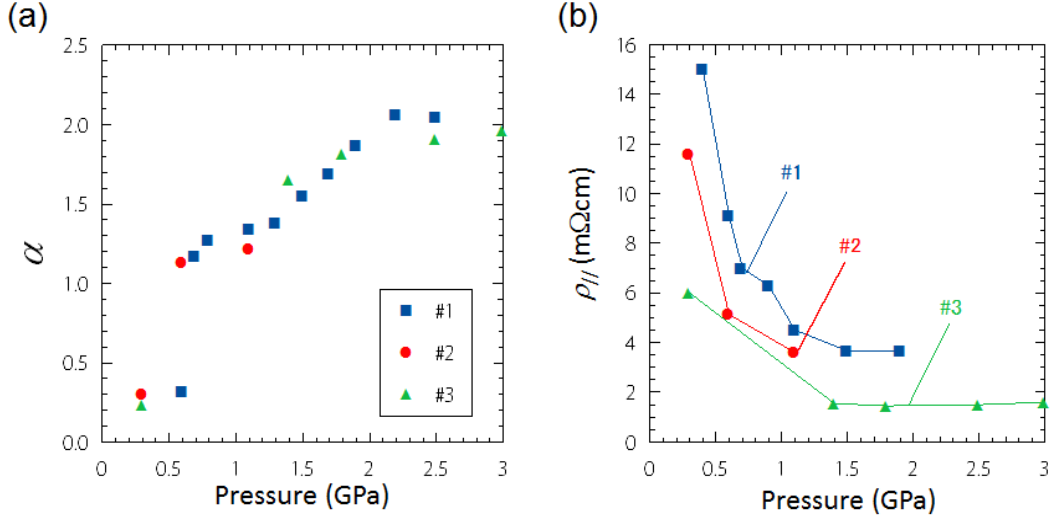

Fig. 6. Reproducibility of contactless conductivity measurement. (a) Pressure dependence of the exponent $\alpha$ in $\rho_{//}-\rho_o \sim T^\alpha$ below 15K. (b) Absolute values of $\rho_{//}$ at 10K. $\alpha$ and absolute values of $\rho_{//}$ shows reproducible pressure dependences.

|  | sample #1 | sample #2 | sample #3 |
|---|---|---|---|
| size of conducting plane | 1.4mm × 1.0mm | 1.2mm × 1.0mm | 1.8mm × 1.2mm |
| thickness of the sample | 0.4mm | 0.5mm | 0.4mm |
| $f_{dip}$ | ~55MHz | ~105MHz | ~85MHz |
| $T_o$ | 100K | 190K | 190K |
| $r$ | 0.6mm | 0.6mm | 0.7mm |

Table 1. Parameters needed in the analyses. The typical dimensions of three samples, $f_{dip}$ values in measurements on each sample, and the parameters, $T_o$ and $r$, which are necessary to the analysis to obtain $\chi_{rf}'$ and $\rho_{//}$.

Although $\rho_{//}$ deviated from each other at high temperatures due to the three origins described above, at low temperatures the pressure dependence of the exponent $\alpha$ in a form of $\rho_{//}-\rho_o \sim T^\alpha$, which fits the data below 15K was reproducible (Fig. 6(a)). In addition, the pressure dependences of $\rho_{//}$ at 10 K show the similar behavior to each other (Fig. 6(b)). The differences of the absolute values by a factor of three are considered to come from the ambiguity in the estimation of the perimeter of the conducting plane and possible sample dependence. Therefore, we regarded the temperature and pressure dependences of $\rho_{//}$ in the small-$\delta/r$ range as reliable.

## II. TEMPERATURE DEPENDENCE OF HALL COEFICIENT

Fig.7 shows temperature dependence of $1/R_H$ under pressure, where $R_H$ is Hall coefficient. $1/R_H$ is approximately temperature independent at 0.85 GPa as expected in the conventional Fermi liquid picture. However, the temperature dependence of $1/R_H$ is remarkable in a low pressure range. The absolute value of $1/R_H$ takes lower in a low pressure range indicating that the effective carrier density decreases by lowering pressure. The decrease of effective carrier density occurs in a metallic state because $1/R_H$ takes a finite value at the lowest temperature in a pressure range where $1/R_H$ drastically changes.

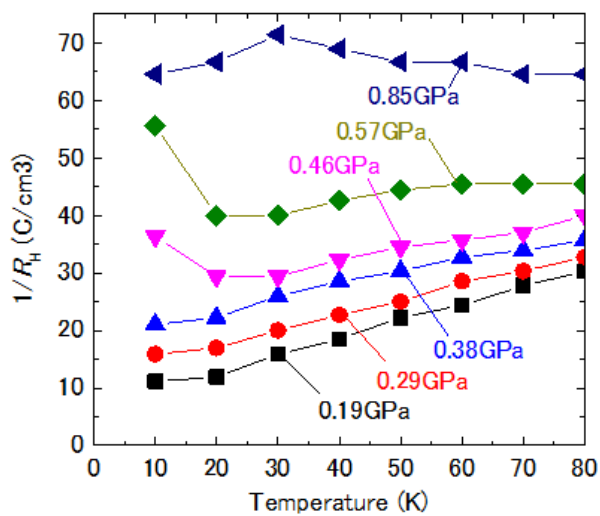

Fig. 7. Temperature dependence of $1/R_H$ under pressure.

**References for supplementary information**

1. H. Oike, K. Miyagawa, K. Kanoda, H. Taniguchi, and K. Murata, Physica B **404**, 376 (2009).
2. K. Kanoda, K. Akiba, K. Suzuki, K. Takahashi, and G. Saito, Phys. Rev. Lett. **65,** 1271 (1990).
3. A. Carrington, I. J. Bonalde, , R. Prozorov, R. W. Gianetta, A. M. Kini, J. Schlueter, H. H. Wang, U. Geiser, and J. M. Williams, Phys. Rev. Lett. **83,** 4172 (1999).
4. S. Milbradt, A. A. Bardin, C. J. S. Truncik, W. A. Huttema, A. C. Jacko, P. L. Burn, S.-C. Lo, B. J. Powell, and D. M. Broun, Phys. Rev. B **88**, 064501 (2013).
5. L. D. Landau, E. M. Lifshitz, and L.P. Pitaevskii, *Electrodynamics of Continuous Media Second Edition* (Pergamon Press, 1984).